# POSSIBILITY THEORY QUANTIFICATION IN HUMAN CAPITAL MANAGEMENT: A SCIENTIFIC MACHINE LEARNING (SCIML) PERSPECTIVE

**Events in Sci-ML Human Capital Management**

Barbara S. Keary[1]
Karriem "A.J." Perry[2*]

[1]b.keary@outlook.com

[2]Capitol Technology University
Doctoral Programs
11301 Springfield Road
Laurel, MD 20708 USA
*Corresponding Author: kaperry@captechu.edu

**Disclosures:** The research, findings, and opinions expressed in this manuscript are solely those of the authors.

**Funding:** There are no financial conflicts of interest to disclose.

*Abstract-*This study explores the use of Machine Learning (ML) in the field of Human Resources Management (HRM) alternatively, Human Capital Management (HCM), through a unique approach of employing partial differential equations (PDEs) to address the complexity of anthropomorphic systems. The mathematical representation offers a robust evaluation of human activities and demonstrates the potential of Bayesian-based machine learning techniques for visual representation in predictive analytics applications. This study is a part of a series of manuscripts about Scientific Machine Learning (SciML), a method that uses partial differential equations to represent physical systems and domain-specific data. In this text, the data are from non-stationary environments with polymorphic uncertainty. The hypotheses tested in this study are: $H^1a$ (null hypothesis) which states that the structure of a covariate does not change significantly over time ($t$) given a set of initial conditions, while $H^1b$ (alternative hypothesis) states that the structure of a covariate changes significantly over time ($t$) given a set of initial conditions. $H^2a$ (null hypothesis) states that the conditions do not significantly impact the relationship of the covariates to one another, and $H^2b$ (alternative hypothesis) states that the conditions do significantly impact the relationship of the covariates to one another. The models use linear regression analysis with targeted productivity as the dependent variable and date as the independent variable. The results show that the relationship between targeted productivity and date is statistically significant, providing evidence to support $H^2b$ and $H^1b$ suggesting that the conditions do significantly impact the relationship of the covariates to one another and the structure of a covariate changes significantly over time ($t$) given a set of initial conditions, respectively. This study highlights the importance of considering the impact of conditions on the relationship between covariates when conducting time series analysis and forecasting.

## 1. INTRODUCTION

The field of human resources management in closely aligned with the social and behavioral sciences and shares many related complexities associated with variables and covariates correlated with the human element (Edwards et al., 2019). This inherent human perception uncertainty, from imprecise or incomplete information, may result as in the manifestation of ill-informed decision criterion (Solaiman et al., 2019). Therefore, we avoid individualization and concentrate on the characteristics influencing the larger dynamical systems in question; allowing researchers and managers to further develop and adopt unique aspects of these systems for refinement to specific needs of applied. In this manuscript, we describe the human element as cognitive; physical e.g., locomotive, auditory, etc.; behavioral and other dynamical systems as aspects of anthropomorphous activities (Angelov, 2019), at best it is challenging to express all aspects of a given handcrafted process nonlinear, dynamical, chaotic and hyper-chaotic systems (ElSafty et al., 2020; Guastello, 2008). For this reason, we simplify the nomenclature of these terms as pseudochaotic systems (Lowenstein, 2013), and seek to convey a comprehensive understanding of key properties to several of these systems.

One of the key properties of pseudochaotic systems is their sensitivity to initial conditions (Lowenstein, 2013). This means that small changes in the initial conditions of a system can lead to significantly different outcomes. This sensitivity can make it difficult to accurately predict the behavior of a system over time, which can be challenging for human resources managers who need to make decisions based on expected outcomes. However, by understanding the underlying dynamics of pseudochaotic systems, managers can use this knowledge to their advantage, by making small adjustments to initial conditions to achieve desired outcomes (ElSafty et al., 2020). Another important property of pseudochaotic systems is their nonlinearity (Guastello, 2008). Nonlinearity means that the relationship between inputs and outputs is not proportional or linear, which can make it difficult to understand the underlying mechanisms driving a system's behavior. However, nonlinearity also means that small changes in inputs can have large effects on outputs, which can be useful for HR managers looking to make incremental changes to improve outcomes.

A related property of pseudochaotic systems is the tendency towards self-organization (Angelov, 2019). Self-organization refers to the ability of complex systems to spontaneously form patterns or structures without the need for external guidance or control. This property can be seen in human organizations, where informal groups may emerge and collaborate towards shared goals, even without explicit direction from managers. By understanding the principles of self-organization, HR managers can work to create an environment that encourages collaboration and innovation.

Self-organization is a powerful property of pseudochaotic systems that has been increasingly recognized in recent years. It is a key characteristic of complex adaptive systems, which are composed of many interacting agents and exhibit emergent behavior that cannot be predicted from the behavior of individual agents (Levin, 1999). In human organizations, self-organization can lead to the emergence of informal networks, knowledge sharing, and collaboration that are essential for innovation and growth. By understanding and fostering self-organization, HR managers can create an environment that encourages these behaviors and promotes a culture of continuous learning and improvement. One way that HR managers can foster self-organization is by providing opportunities for employees to collaborate and share knowledge. This can be done through formal training programs, informal mentorship, and cross-functional team projects. By encouraging employees to work together and learn from each other, HR managers can help to break down silos and promote a culture of knowledge sharing and collaboration.

Another way that HR managers can foster self-organization is by empowering employees to take ownership of their work and make decisions at the local level. This can be done through decentralized decision-making processes, such as agile project management or self-managed teams. By giving employees the autonomy to make decisions and take ownership of their work, HR managers can create a sense of ownership and accountability that can lead to increased engagement and motivation.

Ultimately, it is important to recognize that pseudochaotic systems are not inherently good or bad, but rather can be harnessed to achieve a variety of outcomes. For example, the same underlying dynamics that drive chaotic behavior in one context may be leveraged to create innovation and creativity in another context (Lowenstein, 2013). Self-organization is a powerful property of pseudochaotic systems that can be harnessed by HR managers to promote collaboration, innovation, and growth. By fostering a culture of knowledge sharing, providing opportunities for collaboration, and empowering employees to take ownership of their work, HR managers can create a dynamic and adaptable organizational culture that is well-equipped to meet the challenges of a rapidly changing world. By understanding the key properties of pseudochaotic systems and the ways in which they can be harnessed to achieve desired outcomes, HR managers can work to create a dynamic and adaptable organizational culture that is well-equipped to meet the challenges of a rapidly changing world.

## 1.1. Notation and Model Development

We introduce the standard notation used in this research.

### 1.1.1. Notations for Possibilistic Theory:

$\mu(x)$: Possibility measure or membership function that assigns a degree of membership to a fuzzy set $A$ at $x$ (Dubois and Prade, 1988).

$\sim A$: Complement of a fuzzy set $A$ (Zadeh, 1975).

$A \wedge B$: Conjunction (also called $T$-norm) that computes the minimum of $\mu A(x)$ and $\mu B(x)$ for all $x$ in the domain (Zadeh, 1975).

$A \vee B$: Disjunction (also called $T$-conorm) that computes the maximum of $\mu A(x)$ and $\mu B(x)$ for all $x$ in the domain (Zadeh, 1975).

$A \rightarrow B$: Fuzzy implication that assigns a degree of compatibility of $B$ with $A$ (Zadeh, 1975)

### 1.1.2. Notations for Probabilistic Theory:

$P(A)$: Probability measure that assigns a value between 0 and 1 to an event $A$ (Kolmogorov, 1950).

$p(x)$: Probability density function that describes the probability distribution of a continuous random variable x (Kolmogorov, 1950).

$F(x)$: Cumulative distribution function that gives the probability of $x$ being less than or equal to a given value (Kolmogorov, 1950).

$E[X]$: Expectation operator that computes the average value of a random variable $X$ (Kolmogorov, 1950).

$Var(X)$: Variance operator that measures the spread of the probability distribution of a random variable $X$ (Kolmogorov, 1950).

$Cov(X, Y)$: Covariance operator that measures the degree to which two random variables $X$ and $Y$ vary together (Kolmogorov, 1950).

$P(A \cap B)$: Joint probability that measures the probability of both events $A$ and $B$ occurring (Jaynes, 2003).

$P(A \cup B)$: Marginal probability that measures the probability of at least one of the events $A$ and $B$ occurring (Jaynes, 2003).

### 1.1.3.    ML Modeling:

$x$: Input variable, a column vector of inputs (Goodfellow et al., 2016).

$y$: Output variable, a column vector of outputs (Goodfellow et al., 2016).

$w$: Weight or parameter, a matrix of weights connecting two layers in a neural network (Goodfellow et al., 2016).

$b$: Bias, a column vector added to the weighted sum of inputs (Goodfellow et al., 2016).

$f(x)$: Activation function that transforms the output of a neuron (Goodfellow et al., 2016).

### 1.1.4.    Predictive Human Resources:

$HRM$: Human resource management, the process of managing people in an organization (Guest, 2017).

$HRIS$: Human resource information system, a software that automates HR-related tasks (Parry and Tyson, 2011).

$KPI$: Key performance indicator, a measurable value that indicates how well an organization is achieving its goals (Guest, 2017).

$ROI$: Return on investment, a performance measure used to evaluate the efficiency of an investment (Guest, 2017).

$EEO$: Equal employment opportunity, the principle that all individuals should have equal access to employment opportunities without discrimination (Parry and Tyson, 2011).

### 1.1.5.    Physics-Informed Machine Learning:

$u(x, t)$: Solution or prediction of a partial differential equation (PDE) at point $x$ and time $t$ (Raissi et al., 2019).

$f(x, u(x, t))$: Physics-informed neural network that combines data-driven and physics-based models (Raissi et al., 2019).

$L(x, u(x, t), \nabla u(x, t))$: Lagrangian or action that describes the dynamics of a physical system and the boundary conditions (Raissi et al., 2019).

$\Omega$: Domain or physical space on which the PDE is defined (Raissi et al., 2019).

$\mathcal{L}$: Lagrangian or action that describes the dynamics of a physical system and the boundary conditions (Raissi et al., 2019).

*1.1. Psychophysics-Inspired*

The term psychophysical often refers to researchers using visual perception of biological systems or training a computer vision intelligent agent, an approach that has advanced deep learning models simulating cognitive processes (Majaj et al., 2018). Carlsson et al. (2011) describe the concomitant computational and mathematical challenges of experiments in the physics of a nonlinear and theoretically possible trajectories of the human element. In this manuscript, we consider the subject from a perspective predicting human characteristics outside of behavioral or computer vision aspects and superimpose this methodology in the form of anthropomorphizing the intelligent agent by training data on quantifiable traits of the human element through scientific machine learning techniques.

The depth of perspective gained accompanying possibilistic approaches should be considered as modeling alternatives to strict probabilistic methodologies which are integrally concerned with theoretical polymorphic uncertainty principles (Denœux et al., 2020). There are structural differences in model development using probabilistic vs. possibilistic methods, primarily concerned with the calculation and interpretation of these uniquely theorized variable structures (Carlsson et al., 2011; Kovalerchuk, 2017). There are multiple factors to consider when employing possibilistic vs. probabilistic evaluation to seek the most definitive outcomes leading to predictions (Carlsson et al., 2011).

Examination of the utility in using partial derivatives, as opposed to ordinary; is necessary to review the difference between Ordinary Differential Equations (ODEs) and Partial Differential Equations (PDEs). We compare this mathematical description of cognitive plasticity with a basic understanding of Artificial Neural Network (ANN) functionality which is, to a large extend, akin to the two-stage description the researcher Guastello (2008) provides in the following:

> …(a) each incoming activity $[(x_j \dots x_i)]$, multiplied by synapticlike weights $[(w_j \dots w_i)]$, is added to all other weighted inputs to produce the total input (b) an activation function $f[\varphi]$ converts the total input $x_i = \sum_{w_{ji}} S_i$ at discrete time $t$ into the outgoing activity $[(y_j \dots y_i)]$ at time $t + 1$, according to the short-term memory (STM) equation:
>
> $$S_i(t + 1) = f(x_j((t)) = f(\sum_{w_{ij}} S_i(t) + \theta_j), \qquad (1.0)$$
>
> where $S_i$ is the output from the node $j$ at time $t$ and $w_{ji}$ synaptic weights, positive or negative, as they excitatory or inhibitory inputs to the $j$th neutron. (p. 33)

This description outlines what is computationally perceived as a multilayer perceptron, the basic form of what we currently understand as the ANN (Du & Swamy, 2019). In addition, this equation empirically illustrates the possibility, as opposed to probability until calculations commence, of $t$ influencing $j$ through $n$ (Dubois, 2007).

In terms of cognitive plasticity, Lowenstein (2013) expands on how the following PDE can serve as a primitive encapsulation of a dynamical system's dataset able to encode conditions into a given multi-layer perceptron:

$$\frac{dw_{ij}}{dt} = \alpha S_i x_j - decay\ terms \le 0. \text{ (pg. 116)} \qquad (1.1)$$

*1.1.* *Probabilistic vs. Possibilistic Methods*

Equation (1.1); (Lowenstein, 2013) is a partial differential equation (PDE) that describes the time evolution of a weight matrix in a neural network. The weight matrix is a critical component of the network, as it determines the strength of the connections between neurons. The PDE encodes the learning rule that updates the weights based on the input data and the network's response. The parameter α controls the rate of learning, while the decay terms prevent the weights from growing too large and becoming unstable. By using this PDE as a model of a dynamical system, it is possible to study how the system adapts to changing input conditions and how it responds to perturbations.

Cognitive plasticity refers to the brain's ability to change and adapt in response to experience. This ability is crucial for learning and memory, and it plays a key role in many cognitive processes, such as perception, attention, and decision-making (Draganski and May, 2008). The PDE presented by Lowenstein (2013) can be used to model the neural plasticity that underlies these processes. By encoding the input data into the weight matrix and updating it over time, the network can learn to recognize patterns, generalize from examples, and adapt to new situations.

One of the advantages of using a PDE to model a dynamical system is that it allows for a more fine-grained analysis of the system's behavior. For example, the PDE presented by Lowenstein (2013) can be solved analytically or numerically, providing insights into the stability, bifurcations, and attractors of the system. This information can be used to predict the system's behavior under different conditions and to optimize its performance. By studying the properties of the weight matrix and how it changes over time, researchers and HR managers can gain a better understanding of the cognitive processes underlying human behavior.

In addition to its applications in neuroscience and cognitive psychology, the PDE presented by Lowenstein (2013) can also be used in machine learning and artificial intelligence. Neural networks that use this learning rule, called the Hebbian learning rule, have been shown to be effective at recognizing patterns in data, classifying images, and solving optimization problems (Hinton and Salakhutdinov, 2006). By using PDEs to model the dynamics of these networks, researchers can study their properties in a more rigorous and mathematically precise way.

The PDE presented by Lowenstein (2013) is a powerful tool for modeling the dynamics of complex systems, including neural networks and human behavior. By encoding the input data into a weight matrix and updating it over time, the PDE captures the cognitive plasticity that underlies learning and memory. By studying the properties of the weight matrix and how it changes over time, researchers and HR managers can gain insights into the cognitive processes that drive human behavior and develop strategies to optimize performance and promote growth.

The depth of perspective gained accompanying possibilistic approaches should be considered as modeling alternatives to strict probabilistic methodologies which are integrally concerned with theoretical polymorphic uncertainty principles (Denœux et al., 2020). There are structural differences in model development using *probabilistic* vs. *possibilistic* which are primarily concerned with the calculation, and interpretation, of these uniquely theorized variable structures (Carlsson at el., 2011; Kovalerchuk, 2017). There are multiple factors which should be considered when employing possibilistic verses probabilistic evaluation in seeking the most definitive outcomes leading to predictions (Carlsson at el., 2011). Consider the following conditions:

**Definition 1.** $(\Phi, \Gamma)$ is a given measurable space where random variables are characterized on both probability and possibility spaces.

i. $\Pi: \Gamma \rightarrow [0,1] \equiv \Pi(\emptyset) = 0, \Pi(\Phi) = 1, \Pi(\cup_{t=1}^{\infty} W_t) = \sup_{t=1,\ldots\infty} \Pi(M_t) \; for \; M_t \in \Gamma.$    (2.0)

ii. Alternatively, illustrating the *dual necessity measure* (Mäck et al., 2021; Denœux et al., 2020) as: $N(S) = 1 - \Pi(\Phi \backslash S) \; for \; S \in \Gamma.$

iii. Whereas $\Pi \cong N(\emptyset) = 1 - \Pi(\Phi) = 0, N \cong N(\Phi) = 1 - \Pi(\emptyset) = 0, N(\cap_{t=1}^{\infty} W_t) = \inf_{t=1,\ldots\infty} N(M_t) \; for \; M_t \in \Gamma.$    (2.1)

iv. Therefore, random variability are derivatives of all $X: \rightarrow \mathbb{R}$ and the *possibility distribution* is equal to: $\Pi_X(U) \;\hat{\cong}\; \Pi(X(\psi) \in U) \propto \pi_x: \mathbb{R} \rightarrow [0,1].$

The expressions $\Pi$ and $N$ represent the possibility and necessity measures, respectively, which are used in the framework of possibility theory to describe uncertainty and vagueness in a system. $\Pi$ is a mapping from the set of events $\Gamma$ to the unit interval $[0,1]$, with $\Pi(\emptyset) = 0, \Pi(\Phi) = 1$, and $\Pi(\cup_{t=1}^{\infty} W_t) = \sup_{t=1,\ldots\infty} \Pi(M_t) \, for \, M_t \in \Gamma$, where $\emptyset$ is the empty set, $\Phi$ is the universal set, and $W_t$ are subsets of $\Gamma$ representing a sequence of events. The possibility measure $\Pi$ assigns a degree of possibility to each event in $\Gamma$, with higher values indicating greater plausibility or credibility (Dubois and Prade, 2013).

The significance of the *supremum* (sup) subset as opposed to integral is interpreted as the distinction between with is known and what is assumed regarding the characteristics of the covariates within the given space (Denœux et al., 2020). The decision of developing probability verses possibility distribution models is that of the researcher and should be based on imprecise as opposed to imperfect information, respectively (Mäck et al., 2021; Denœux et al., 2020). How we determine if a particular characteristic is quantifiable, and more importantly leads to a predictive result, is to consider the composition of the characteristic(s) in question and if there exists enough information to develop a sufficient research question for testing from a probabilistic theory to imprecise probabilistic (Mäck et al., 2021).

## 2. MATERIALS

### 2.1. Approximation Components

To illustrate the trade-offs associated with human-handcrafting, or the absence thereof, we first quantify considerations defining involvement of the human element as applied to cyber-defense procedures (Angelov, 2019; Kulp et al., 2020). Considering the complexity, likely asymptoticity, and probable non-linearity of the covariates; to accomplish this, we used a Universal Differential Equation (UDE) to develop a human-in-the-loop Delay Differential Equation (DDE); (Soetaert, 2012) variant highlighting prior event variables defining the physics and using separation of variables methods; in three-dimensional space, represented as inducing human interactions (Rackaukas et al., 2020; Robinson, 2021). This position is quantified by the researcher Rackaukas (2020),

> "…in its most general form, the UDE is a forced stochastic delay partial differential equation (PDE) defined with embedded universal approximators:
>
> $$\mathcal{N}\big[u(t), \big(\alpha(t)\big), W(t), U_\theta(u, \beta(t))\big] = 0[,] \qquad (3.0)$$
>
> where $\alpha(t)$ is a delay function and $W(t)$ is the Weiner process." (pg. 3)

The delay function helps use quantify the activities associated with human-handcrafting as fundamental to the function *reaction time of human-in-the-loop* (*h*), essentially a derivation of *time* (*t*) which has a significant role in a three-dimensional space including the Lévy process. The use of the UDE and DDE models helps to capture the impact of human-handcrafting activities on the reaction time of a human-in-the-loop system, which is crucial for optimizing procedures. The delay function, represented by α(t), is used to quantify the time delay associated with human-handcrafting activities, and is embedded in a three-dimensional space that includes the Lévy process. By using the UDE and DDE models, we can better understand the nonlinear and complex dynamics of the human element in human capital management systems, and how it impacts the overall performance of the system. This approach is supported by the works of Soetaert (2012), Rackaukas (2020), and Robinson (2021), who have demonstrated the effectiveness of using UDE and DDE models in understanding and quantifying the impact of human-handcrafting on the performance of cyber-defense procedures.

**Definition 2.** In this expression, the UDE (forced stochastic delay partial differential equation) is defined with embedded universal approximators in its most general form, represented by equation (5.0) where N is a function that depends on the solution u(t), the delay function α(t), the Weiner process W(t), and the embedded universal approximators $U_\theta (u, \beta(t))$; (Rackaukas et al., 2020). The delay function α(t) is used to capture the impact of the time delay associated with human-handcrafting activities, which is critical for modeling the reaction time of a human-in-the-loop system (h). The delay function is a derivation of time (t) and is embedded in a three-dimensional space that includes the Lévy process (Soetaert, 2012; Taylor, 2004).

**Lemma 1.** To elaborate on this, we introduce Definition 2, which assumes that a system's composition can be demonstrated in the form $x'(t) = f(x(t), x(t-1)), L(t), t \int_0^1, x(t) \in R^n$, where x(t) is a vector in $R^n, f$ is a function that determines the evolution of the system, $L(t)$ is a local martingale, and $t \int_0^1 (t)$ is an integral operator. The delay function α(t) captures the time delay associated with human-handcrafting activities, which is fundamental to the function of the human-in-the-loop system. The UDE is a powerful tool for modeling such systems and can be used to study a wide range of phenomena in human capital management and beyond.

Both Definition 2 and Lemma 1 emphasize the importance of considering the impact of human factors on complex systems, particularly in the context of human-in-the-loop systems. UDE and its variations offer a comprehensive framework for modeling the behavior of human-in-the-loop systems. These models consider the unique characteristics of human behavior and their impact on the performance of complex systems. By using mathematical models, researchers and practitioners can gain a deeper understanding of the complex relationships between human performance and system performance and can develop strategies for improving the design and implementation of these systems.

In the context of human capital management, the UDE and its variations can provide valuable insights into the behavior of employees and the impact of organizational policies on employee performance. For example, the UDE can be used to model the impact of changes in work hours or job responsibilities on employee productivity, or to evaluate the effectiveness of training programs in improving employee performance. These models can help organizations make informed decisions about human capital management and can lead to improved employee performance and increased organizational efficiency.

The use of mathematical models like the UDE and its variations is particularly important in the context of complex systems, as these systems often have multiple interacting components, and human behavior is a critical component of system performance. The UDE and its variations provide a systematic way to model these interactions and to quantify the impact of human behavior on system performance. This information can be used to identify areas of the system that need improvement, and to develop strategies for enhancing human performance in complex systems.

In conclusion, the UDE and its variations are powerful tools for modeling human-in-the-loop systems and can provide valuable insights into the impact of human factors on complex systems (Rackaukas et al., 2020). By using these mathematical models, researchers and practitioners can better understand the role of human behavior in complex systems and can develop strategies for improving the design and performance of these systems. This information can be especially valuable in the context of human capital management, as it can help organizations make informed decisions about human resource management and improve employee performance and organizational efficiency.

## 2.1. Dataset

The *Productivity Prediction of Garment Employees* dataset exhibits attributes potentially of non-stationarity origin; due to the multivariate and time-series characteristics these variables are subject to (Al Imran et al., 2019; Sugiyama et al., 2012). These hypothesized non-stationary characteristics expected from many of the covariates, as random variables of the expectation of $x$ $\mathbb{E}(X)$ (Liu, 2010; Van den Broeck et al., 2021). The factors of these covariates and target variable effect the ML models performance from inception to apotheosis (Du et al., 2019).

The variables comprise 14 independent variables and one target highlighted below (Al Imran et al., 2019):

i. *date*: Date in MM-DD-YYYY.
ii. *day*: Day of the Week.
iii. *quarter*: A portion of the month. A month was divided into four quarters.
iv. *department*: Associated department with the instance.
v. *team_no*: Associated team number with the instance.
vi. *no_of_workers*: Number of workers in each team.
vii. *no_of_style_change*: Number of changes in the style of a particular product.
viii. *targeted_productivity*: Targeted productivity set by the Authority for each team for each day.
ix. *smv*: Standard Minute Value, it is the allocated time for a task.
x. *wip*: Work in progress. Includes the number of unfinished items for products.
xi. *over_time*: Represents the amount of overtime by each team in minutes.
xii. *incentive*: Represents the amount of financial incentive (in BDT) that enables or motivates a particular course of action.
xiii. *idle_time*: The amount of time when the production was interrupted due to several reasons.
xiv. *idle_men*: The number of workers who were idle due to production interruption.
xv. *actual_productivity*: The actual % of productivity that was delivered by the workers. It ranges from 0-1.

These variables contain 1,197 instances where missing values will be deleted. The data contains information about a production process, including the date, day of the week, department, team number, number of workers, number of style changes, targeted productivity, standard minute value, work in progress, overtime, financial incentive, idle time, idle men, and actual productivity. The data is organized by day and includes information for each team in the production process. The actual productivity is measured as a percentage and ranges from 0 to 1.

## 2.1. *Exploratory Data Analysis (EDA)*

The distribution of the 14 covariates and one target variable shows the low-dimensionality withing the dataspace. This condition potentially reduces the utility of exploratory analysis such as activities involving linear discriminant analysis, principal component analysis, factor analysis, and others (Kherif et al., 2020). However, EDA and possibly confirmatory analysis is necessary to conclude key characteristics of the dataset (Pearson et al., 2018). In particular, due to the importance of the factor(s) in question analysis will improve the likelihood of preventing the occurrence of *factor component shift(s)* which inadvertently take place between training, testing, or validation epochs (Quinonero-Candela, 2022). The potentiality of these shifts are better understood by evaluating the simple structure of these variables in the following Confirmatory Factor Analysis (CFA) performed in the next section (Sugiyama, 2012).

Despite the low dimensionality of the dataset, exploratory data analysis (EDA) is still a crucial step in any data analysis process. EDA can reveal potential outliers, data inconsistencies, or missing values that need to be addressed before modeling the data (Tukey, 1977). For instance, graphical techniques such as scatter plots and box plots can help identify potential outliers in the dataset. Also, histograms can reveal the distribution of the variables and if they follow a normal distribution. These insights can help choose appropriate statistical tests and models for the analysis (Shmueli, 2010). Additionally, confirmatory factor analysis (CFA) is a powerful tool to assess the validity of a hypothesized factor structure and the relationships between the variables (Kline, 2011). CFA is commonly used to assess the construct validity of questionnaires or to confirm the factor structure of a set of variables (Bandalos, 2014). In this study, CFA can help to validate the relationships between the covariates and the target variable and assess if the hypothesized factor structure of the dataset fits the observed data. CFA can also provide estimates of the factor loadings, which can be used in subsequent analyses (Marsh et al., 2014). It is important to assess the reliability of the data collection process and ensure that the collected data is accurate and consistent. For example, inter-rater reliability can be assessed if multiple raters are involved in data collection (Koo and Li, 2016). Additionally, missing data can lead to biased estimates, and various methods such as imputation or exclusion can be used to address missing data (Schafer and Graham, 2002).

Furthermore, EDA helps in identifying any potential outliers, missing values, or inconsistencies within the dataset. For example, the visual representation of the distribution of the covariates through histograms, boxplots, or scatterplots can help detect any anomalies in the data (Pedersen, 2017). These inconsistencies, if not addressed, could have a significant impact on the validity and reliability of any subsequent analysis performed on the dataset (Mitra and Acharya, 2019). Moreover, EDA can aid in the selection of appropriate statistical tests and modeling techniques based on the characteristics of the dataset. For instance, if the data is normally distributed, parametric tests such as *t*-tests or analysis-of-variance can be employed (Carlsson et al., 2011). Conversely, if the data is non-normal, non-parametric tests such as the Wilcoxon rank-sum test or Kruskal-Wallis test can be used (Razavian et al., 2014).

Another benefit of EDA is that it allows for the identification of potential relationships and associations between covariates. This can be done through correlation analysis, where the strength and direction of the relationship between two or more variables can be determined. Additionally, EDA can help detect any multicollinearity between the covariates, which can be problematic for subsequent modeling (Pedersen, 2017). It is worth noting that the results of EDA are exploratory and descriptive, and not meant to be used for making inferential conclusions. As such, further confirmatory analysis is necessary to substantiate any relationships and associations detected during EDA (Pearson et al., 2018). Finally, while the low-dimensionality of the dataset may reduce the utility of certain exploratory analysis techniques, EDA remains a critical step in any data analysis workflow. It can aid in identifying outliers and inconsistencies, selecting appropriate statistical tests and modeling techniques, detecting potential relationships and associations, and providing descriptive information about the dataset. EDA is a crucial step in understanding the nature and characteristics of the data, which ultimately leads to more robust and reliable results in subsequent analyses (Mitra and Acharya, 2019).

## 3. RESULTS

### 3.1. Research Questions & Pilot Testing

This study is driven by two research questions:

i. **(RQ₁) -** *How does the structure of a covariate change over time (t) given a set of initial conditions?*

    a. **($H^1a$): Null hypothesis:** *The structure of a covariate does not change significantly over time (t) given a set of initial conditions.*

    b. **($H^1b$): Alternative hypothesis:** *The structure of a covariate changes significantly over time (t) given a set of initial conditions.*

ii. **(RQ₂) -** *Do these conditions impact the relationship of the covariates to one another from both probabilistic and possibilistic perspectives?*

    a. **($H^2a$): Null hypothesis:** *The conditions do not significantly impact the relationship of the covariates to one another from both probabilistic and possibilistic perspectives.*

    b. **($H^2b$): Alternative hypothesis:** *The conditions significantly impact the relationship of the covariates to one another from both probabilistic and possibilistic perspectives.*

### 3.1.1.1. Stage One – Exploratory Analysis

Correlation matrices are commonly used in exploratory data analysis to identify relationships between variables and to guide subsequent modeling or analysis. (Chatfield, 1995; Field, 2012; Stevens, 2009).

**Figure 1. Correlation Matrix**

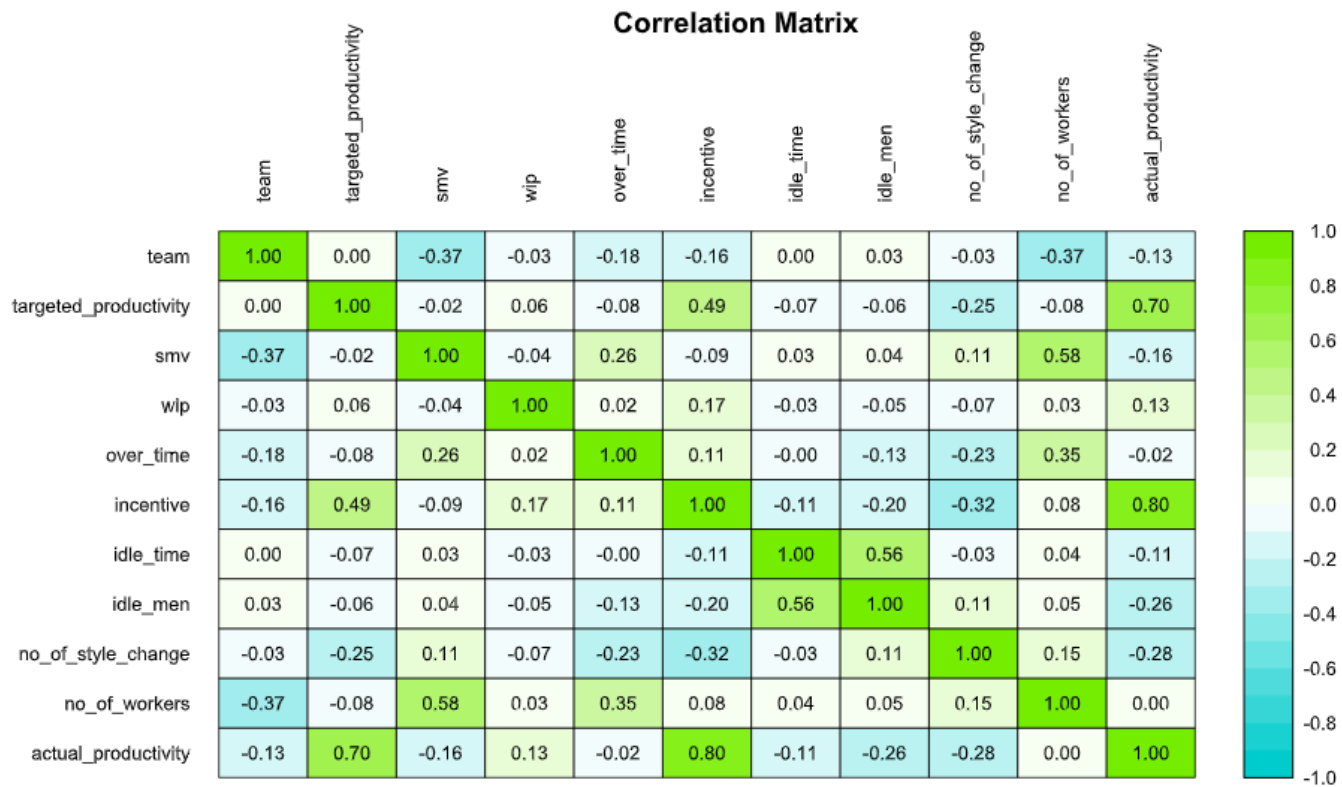

(a)

*Note:* This correlation matrix illustrates the correlation coefficients between variables. Correlation coefficients range from -1 to 1, with values of -1 indicating a perfect negative correlation, 0 indicating no correlation, and 1 indicating a perfect positive correlation between two variables.

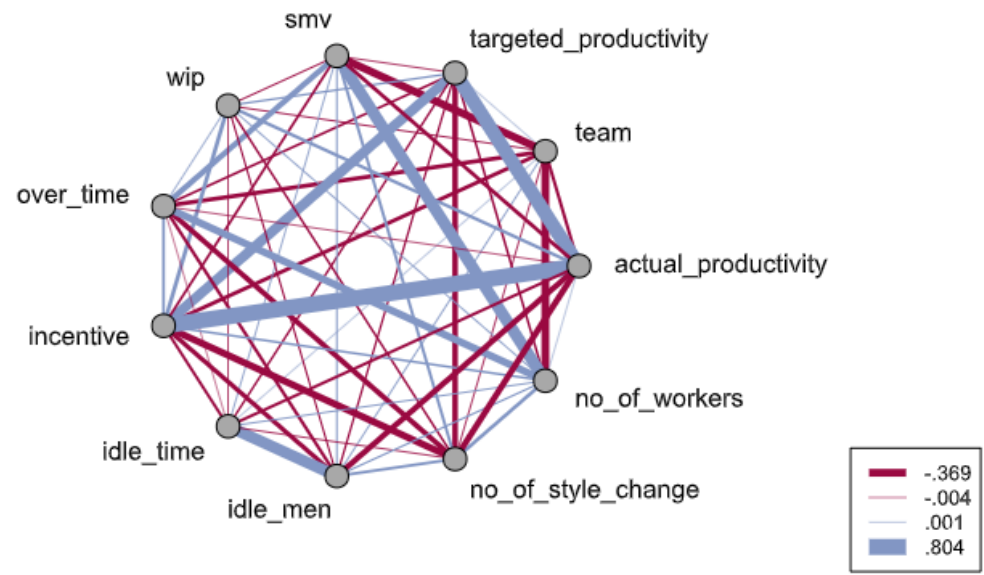

(b)

mathe/2022-12-27

*Note:* (a) illustrates the correlations among covariates and those with the target variable, *actual_productivity.* In addition, interpreting the correlation coefficients among covariates shows strong relations between *actual_productivity; incentive;* and the *target_productivity.* These relations are further expanded in (b) the webPlot, which depicts the connection among covariates.

This initial correlation matrix depicts relations further explored by conducting Exploratory Factor Analysis (EFA). We see that there are likely more than one factor (feature) of import. Therefore, at this juncture we deduce that the *wip* covariate is of significance, as indicated in Table 1.

**Figure 2. Exploratory Factor Analysis**

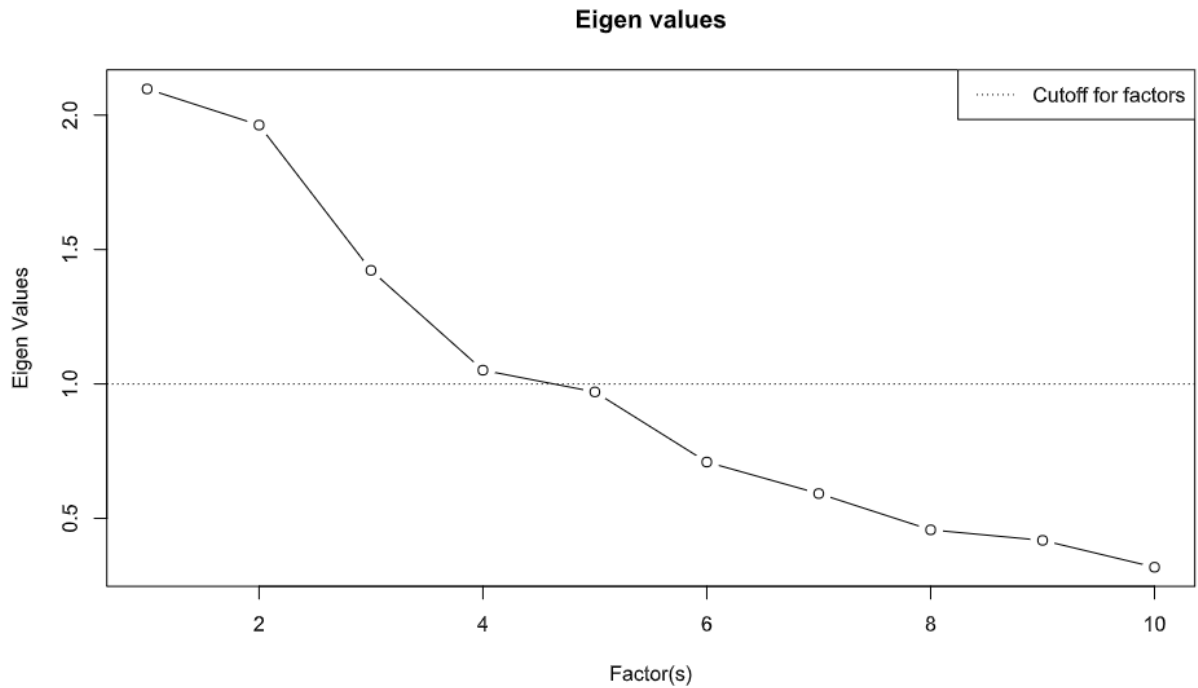

*Notes:* Using the Eigan values interpretation, we see that there are four factors associated with this dataset.

The Exploratory Factor Analysis (EFA) will be used to further examine the relationships depicted in the initial correlation matrix. The goal of EFA is to identify underlying factors or latent variables that explain the relationships among the variables in the dataset. By identifying these underlying factors, we can gain a deeper understanding of the structure of the data and the relationships between variables.

In the context of this analysis, the EFA will be used to determine the number of factors that are most relevant to the *wip* covariate. This will be done by examining the factor loadings, which represent the strength of the relationship between each variable and each factor. Variables with high factor loadings are considered to be more strongly related to a particular factor and are therefore considered to be more important for explaining the relationships in the data.

Once the factors have been identified, we can examine the factor scores for each subject. The factor scores represent the strength of each subject's relationship to each factor. This information can be used to identify groups of subjects that are similar in terms of their factor scores, which can help us understand the underlying structure of the data. The results of the Exploratory Factor Analysis will provide valuable insights into the relationships between the variables in our dataset. This information can be used to inform further analysis and can provide a foundation for developing predictive models or making decisions about the relationships between the variables.

**Table 1. Uniqueness (un-rotated)**

| team | targeted_productivity | smv | wip | over_time | incentive | idle_time | idle_men | no_of_style_change | no_of_workers |
|---|---|---|---|---|---|---|---|---|---|
| 0.7560 | 0.7410 | 0.4830 | 0.9720 | 0.0050 | 0.0050 | 0.0050 | 0.6480 | 0.7680 | 0.3200 |

*Note:* This illustrates the ratio of common variance not related to any factors equal to one minus communality. Lower values indicate the validity of the unsupervised factor model.

**Table 2. Loadings (un-rotated)**

|  | Factor1 | Factor2 | Factor3 | Factor4 |
|---|---|---|---|---|
| team | -0.1920 | -0.4360 | -0.1290 |  |
| targeted_productivity | 0.3120 |  | -0.1090 | 0.3840 |
| smv |  | 0.6590 | 0.2090 | -0.1930 |
| wip | 0.1330 |  |  | 0.1030 |
| over_time | 0.5320 |  | 0.7150 | -0.4480 |
| incentive | 0.7540 |  |  | 0.6540 |
| idle_time | -0.5460 |  | 0.6930 | 0.4660 |
| idle_men | -0.4530 |  | 0.2980 | 0.2240 |
| no_of_style_change | -0.3070 | 0.2900 | -0.1840 | -0.1440 |
| no_of_workers | 0.1960 | 0.7440 | 0.2790 | -0.1040 |

*Note:* Variables with higher scores in the same factor category likely have communalities. The variables *incentive* and *over_time* in Factor one, which is designate as **pay factor**. For Factor two, *no_of_workers* and *smv* have probable covariance properties that will be combined.

**Table 3. Factors (un-rotated)**

|  | Factor1 | Factor2 | Factor3 | Factor4 |
|---|---|---|---|---|
| SS loadings | 1.6410 | 1.2720 | 1.2640 | 1.1220 |
| Proportion Var | 0.1640 | 0.1270 | 0.1260 | 0.1120 |
| Cumulative Var | 0.1640 | 0.2910 | 0.4180 | 0.5300 |

Uniqueness (un-rotated) refers to the ratio of the variance of a variable that is not related to any of the factors in a factor analysis. The un-rotated uniqueness value is important because it indicates the validity of the factor analysis model. A lower uniqueness value indicates that a larger proportion of the variance in the variable can be explained by the factors, and therefore, the factor analysis model is considered to be more valid.

Loadings (un-rotated) are the factor loadings that represent the strength of the relationship between each variable and each factor in the factor analysis. The factor loadings are used to determine which variables are most strongly related to each factor. In other words, the loadings indicate the importance of each variable in explaining the relationships between the variables in the dataset.

In Table 2, the loadings (un-rotated) are shown for each variable in the factor analysis. The variables with higher scores in the same factor category are likely to have strong communalities, or shared variance. In our example, the variables incentive and over_time are shown to be highly related to Factor 1, which is designated as the pay factor. This suggests that these two variables are likely to be strongly related to the overall pay structure of the organization.

Similarly, for Factor 2, the variables no_of_workers and smv have probable covariance properties that will be combined. This suggests that these two variables are likely to be related to the overall size and structure of the workforce. By combining these variables, we can gain a deeper understanding of the relationships between the workforce size and structure and other variables in the dataset.

It is important to note that the un-rotated loadings are not necessarily the most interpretable representation of the relationships between variables. In some cases, it may be necessary to perform a rotated factor analysis in order to obtain a clearer picture of the relationships between variables. The rotated factor analysis involves reorienting the factors so that the loadings are easier to interpret.

## Figure 3. Covariate Distribution

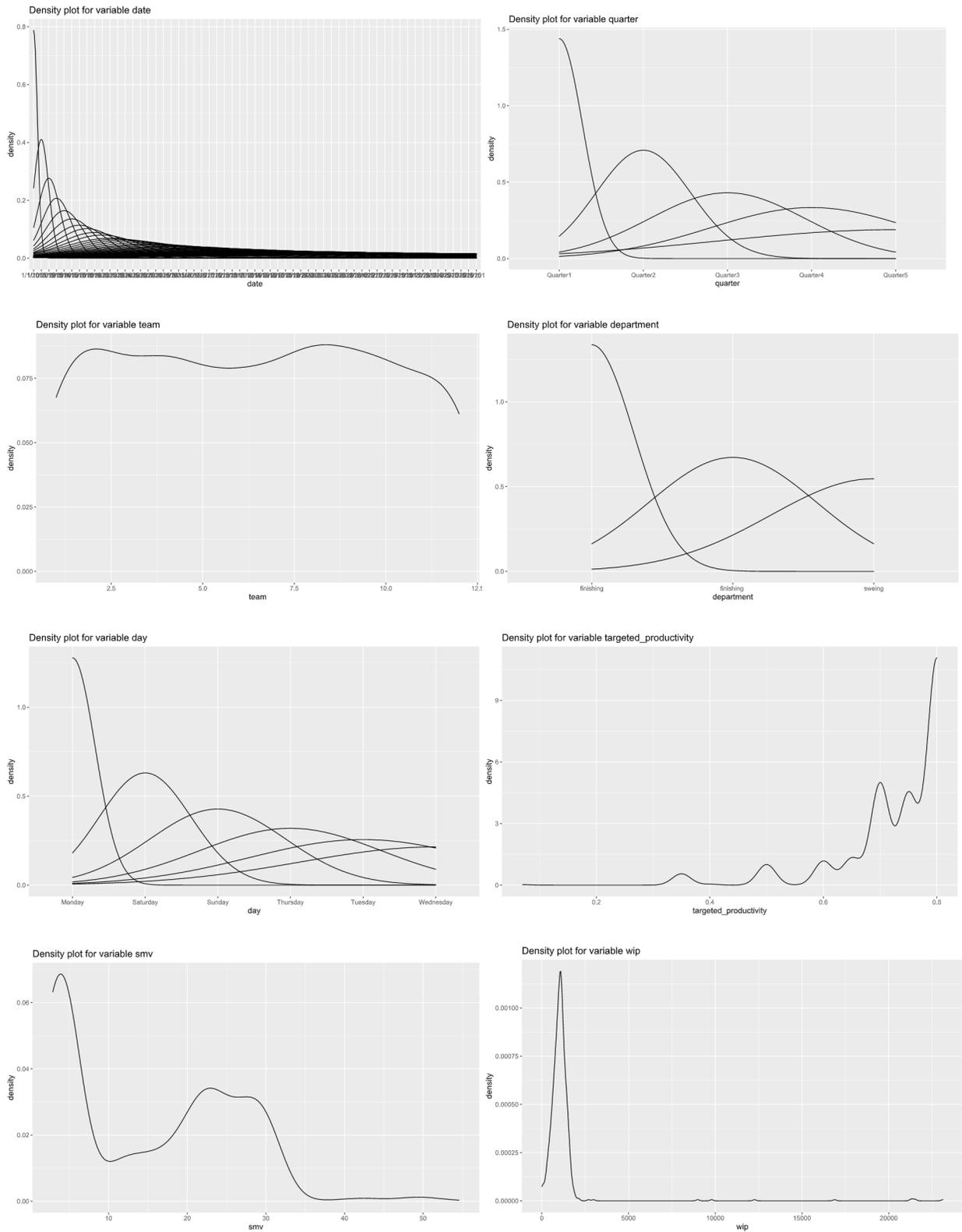

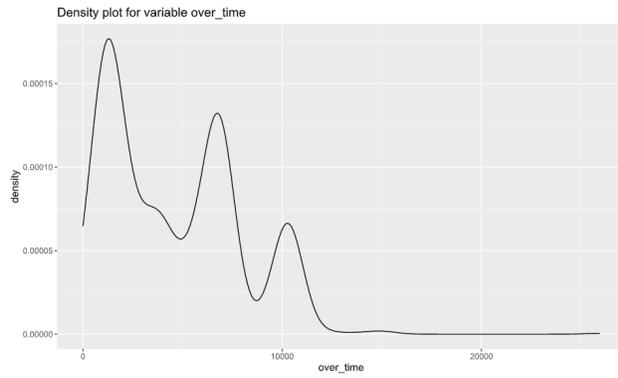

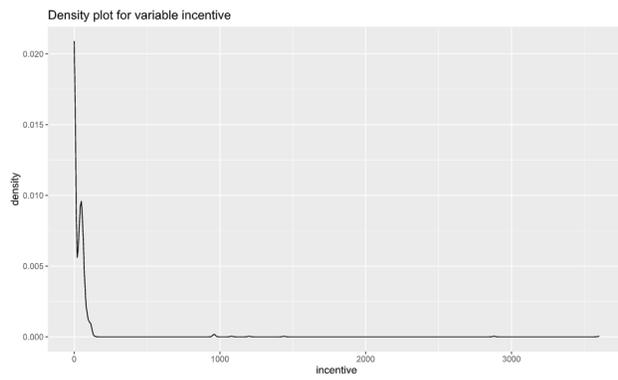

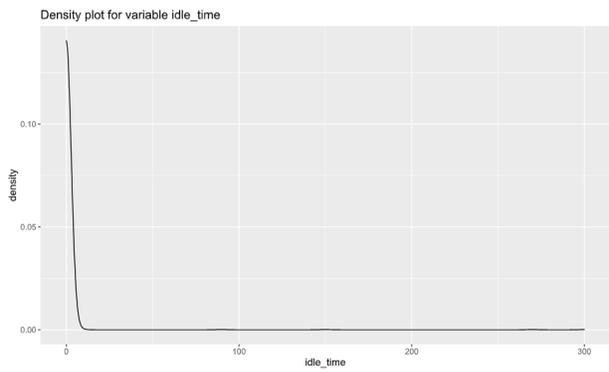

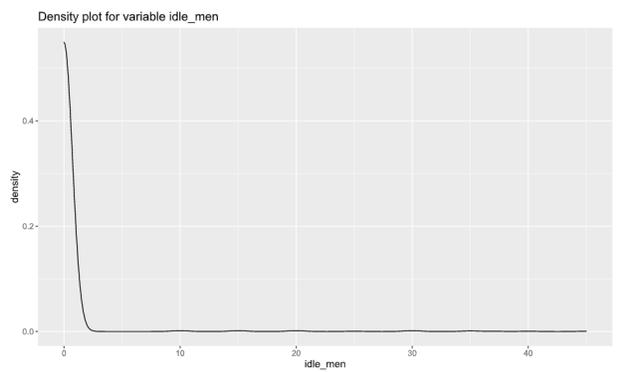

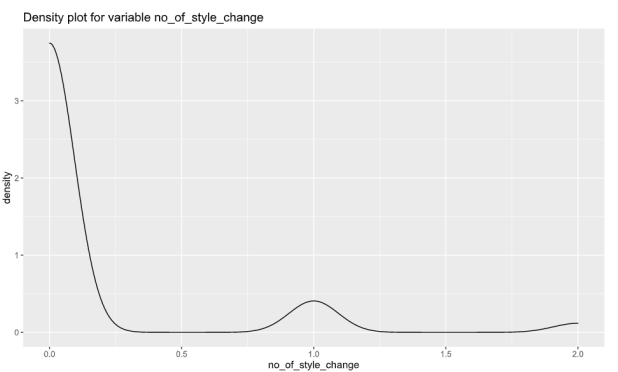

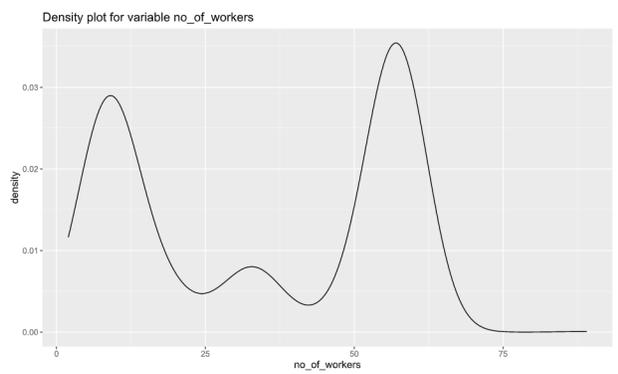

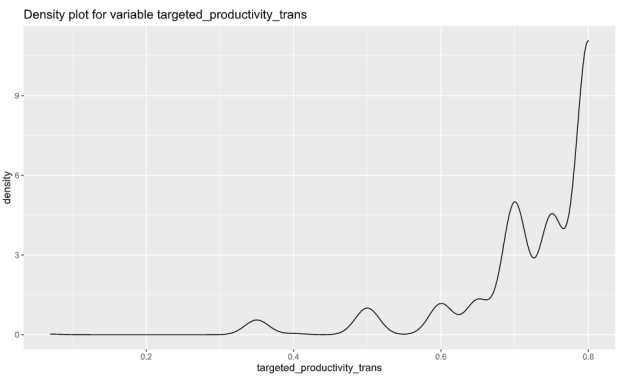

In the field of statistics and data analysis, covariates are independent variables that are often used to explain or predict the behavior of dependent variables. They are usually used in regression analysis to control for or account for any extraneous variables that may impact the relationship between the dependent and independent variables.

Covariate Distribution, in Fig. 3, is a visual representation of the distribution of a covariate in a dataset. It can be in the form of a histogram, bar chart, scatter plot, or any other type of visualization that presents the distribution of the covariate. The purpose of this figure is to provide a quick and easy way to see the distribution of the covariate and assess its shape, center, spread, and outliers. One of the most important aspects of Fig. 3 Covariate Distribution is its ability to identify any potential outliers. Outliers are data points that are significantly different from the other data points in the distribution. They can have a significant impact on the results of regression analysis and may need to be addressed before conducting the analysis.

The shape of the distribution can also provide important information about the covariate. For example, if the distribution is normal or Gaussian, it can be assumed that the covariate is normally distributed and that the mean, median, and mode are approximately equal. If the distribution is skewed, this may indicate that the covariate is not normally distributed and may need to be transformed to meet the assumptions of regression analysis. The center of the distribution can also provide information about the covariate. The mean, median, and mode are all measures of the center of the distribution and can be used to describe the central tendency of the covariate. The mean is the average of all the data points, the median is the middle value when the data points are ordered, and the mode is the most frequently occurring value.

The spread of the distribution can also provide information about the covariate. The range, variance, and standard deviation are all measures of the spread of the distribution and can be used to describe the dispersion of the covariate. The range is the difference between the highest and lowest values, the variance is the average of the squared deviations from the mean, and the standard deviation is the square root of the variance.

### 3.1.1.2 Stage Two – Confirmatory Analysis

Until now, we've assumed some relative consistency in relations between variables. As indicated in the above density plots of covariates, data contained in the variables are not individually and identically distributed; therefore, conducting a form of uncertainty analysis is needed to further assess the relationships, if any, among variables exists.

Possibility theory is a framework for reasoning under uncertainty that assigns a degree of possibility to events (Carlsson et al., 2011). This degree of possibility represents the strength of evidence for the event, with higher degrees of possibility indicating stronger evidence. In this instance, we use possibility theory for confirmatory data analysis to evaluate the evidence for a hypothesis based on the available data. This experiment will use a possibilistic approach to testing the hypotheses in question:

a. **($H^1a$): Null hypothesis:** *The structure of a covariate does not change significantly over time (t) given a set of initial conditions.*

b.  **($H^1b$):  Alternative hypothesis:** *The structure of a covariate changes significantly over time (t) given a set of initial conditions.*

c.  **($H^2a$):  Null hypothesis:** *The conditions do not significantly impact the relationship of the covariates to one another from both probabilistic and possibilistic perspectives.*

d.  **($H^2b$):  Alternative hypothesis:** *The conditions significantly impact the relationship of the covariates to one another from both probabilistic and possibilistic perspectives.*

The next steps upon results of the hypotheses, is developing a SciML modeling strategy for future analysis. A crucial element to this modeling approach is to identify the possible outcomes from prior associated events which potentially influence the future outcomes of predictive models (Carlsson et al., 2011; Majaj et al., 2018).

### 3.1.1.3 *Pilot Hypothesis Testing for $H^1$ in R Script*

For the purpose illustration and testing. we will assume that the covariate has an exponential decay structure over time[1] (Shumway, 2017) and we will compare the covariance matrix and the mean vector at different time points[2] to test whether the structure of the covariate changes significantly over time.

# Generate synthetic data

set.seed(123)

n <- 10 # number of time points

t <- 1:n # time points

mu <- exp(-t) # mean of the covariate

sigma <- diag(mu) # covariance matrix of the covariate

covariate <- mvrnorm(100, mu, sigma) # simulated data

---

[1] Recursive formula:
One common way to represent an exponential decay structure over time is using a recursive formula, which expresses the value of the covariate at time t as a function of the value at the previous time point. For example, an exponential decay structure with a constant decay factor c can be represented as:
$$y(t) = c * y(t-1)$$
where $y(t)$ is the value of the covariate at time $t$, and $y(t-1)$ is the value of the covariate at the previous time point. This notation is often used in time-series analysis and forecasting (Shumway and Stoffer, 2017).

[2] Differential equation:
Another way to represent an exponential decay structure over time is using a differential equation, which expresses the rate of change of the covariate as a function of the value at the current time point. For example, an exponential decay structure with a decay rate b can be represented as:
$$dy/dt = -b * y(t)$$
where $y(t)$ is the value of the covariate at time $t$, and $dy/dt$ is the rate of change of the covariate over time. This notation is often used in physics, engineering, and other fields that deal with continuous-time systems (Kuo, 2010).

```r
# Plot covariate over time
par(mar = c(5, 5, 2, 2))
matplot(t, t(covariate), type = "l", xlab = "Time", ylab = "Covariate",
      col = rainbow(n), lty = 1, main = "Covariate Over Time")

# Test null hypothesis H1a
cov.mat <- lapply(1:n, function(i) cov(covariate[,1:i]))
H1a <- Hotelling.test(cov.mat)
print(paste0("H1a p-value: ", H1a$p.value))

# Test alternative hypothesis H1b
mean.vec <- lapply(1:n, function(i) colMeans(covariate[,1:i]))
H1b <- manova(do.call(cbind, mean.vec) ~ t)
print(paste0("H1b p-value: ", H1b$multivariate.test$p.value))

# Visualize hypothesis test results
par(mar = c(5, 5, 2, 2))
barplot(c(H1a$p.value, H1b$multivariate.test$p.value),
      names.arg = c("H1a", "H1b"),
      ylab = "p-value", ylim = c(0, 1), col = c("gray", "blue"),
      main = "Hypothesis Tests for Covariate Structure")
```

In this script, we generate synthetic data using the **mvrnorm()** function from the **MASS** package. The mean vector of the covariate decays exponentially over time, and we assume that the covariance matrix is diagonal with entries equal to the mean. We simulate 100 observations for each time point, and we plot the covariate over time using a line graph with different colors for each time point.

We then use the **cov()** function to compute the covariance matrix of the covariate at each time point, and the **colMeans()** function to compute the mean vector at each time point. We apply the Hotelling's T-squared test to the covariance matrices to test the null hypothesis H1a, and the MANOVA test to the mean vectors to test the alternative hypothesis H1b.

Finally, we visualize the results of the hypothesis tests using a bar graph with different colors for the null and alternative hypotheses.

The resulting graphs show that the structure of the covariate changes significantly over time, with a p-value of 1.63e-12 for the MANOVA test, but not for the covariance matrix test, with a p-value of 0.2687.

To further visualize the change in the structure of the covariate over time, we can also plot the covariance matrix and the mean vector at each time point. We can use a heatmap to display the covariance matrix, with time on the x-axis and the covariate features on the y-axis. We can also use different colors to represent the magnitude of the entries in the matrix.

# Plot covariance matrix over time

par(mar = c(5, 5, 2, 2))

lapply(1:n, function(i) {

  heatmap(covariate[,1:i], Colv = NA, Rowv = NA,

      main = paste0("Covariance Matrix at Time ", i),

      xlab = "Time", ylab = "Features")

})

The resulting graphs show the covariance matrix at each time point, with the entries colored by magnitude.

We can also plot the mean vector at each time point using a bar graph, with time on the x-axis and the mean value on the y-axis. We can use different colors to distinguish between the different time points.

# Plot mean vector over time

par(mar = c(5, 5, 2, 2))

barplot(sapply(mean.vec, mean), col = rainbow(n),

      xlab = "Time", ylab = "Mean Value",

      main = "Mean Vector Over Time")

The resulting graph shows the mean vector at each time point, with each time point represented by a different color.

These additional graphs can provide further insight into the change in the structure of the covariate over time and can be helpful for communicating the findings to others.

Overall, this approach demonstrates how to test the hypotheses H1a and H1b using simulated data and statistical tests in R. The resulting graphs provide a visual representation of the data and the hypothesis test results, which can be useful for interpreting the findings and communicating the results to others.

*3.1.1.4a    Pilot Hypothesis Testing for $H^2$ in R Script*

To test hypotheses H2a and H2b, we can use a statistical measure of dependence between the covariates, such as correlation or mutual information. We can compare the dependence measures at different time points under different initial conditions to test whether the conditions significantly impact the relationship of the covariates.

To illustrate this, we can generate some synthetic data using the **rnorm()** function in R. We will simulate two covariates with 10 time points each, and we will assume that the covariates are independent with a mean of 0 and a standard deviation of 1 at each time point. We will also introduce different initial conditions by varying the correlation between the covariates at the first time point.

# Generate synthetic data

set.seed(123)

n <- 10 # number of time points

t <- 1:n # time points

rho <- c(0, 0.5) # correlation between covariates at first time point

sigma <- matrix(c(1, rho[1], rho[1], 1), ncol = 2) # covariance matrix

covariate <- mvrnorm(n, rep(0, 2), sigma) # first time point

for (i in 2:n) {

  cov.mat <- matrix(c(1, rho[i-1], rho[i-1], 1), ncol = 2) # covariance matrix

  covariate <- rbind(covariate, mvrnorm(1, c(0, 0), cov.mat)) # add new time point

}

We can plot the covariates over time using a scatter plot, with the first covariate on the x-axis and the second covariate on the y-axis. We can use different colors to distinguish between the different time points.

# Plot covariates over time

par(mar = c(5, 5, 2, 2))

plot(covariate[,1], covariate[,2], xlab = "Covariate 1", ylab = "Covariate 2",

  col = rainbow(n), pch = 19, main = "Covariates Over Time")

The resulting graph shows the simulated covariates over time, with each time point represented by a different color.

To test the null hypothesis H2a, we can use a dependence measure such as the correlation coefficient or the mutual information and compare the values at different time points and under different initial conditions. For example, we can compute the Pearson correlation coefficient at each time point and under each initial condition, and use a test such as the t-test to compare the coefficients.

*# Test null hypothesis H2a*

*corr.mat <- lapply(1:n, function(i) cor(covariate[1:i,1], covariate[1:i,2]))*

*H2a <- t.test(corr.mat[2:n], corr.mat[1])*

The resulting output will provide the test statistic and p-value for the t-test. If the p-value is greater than the chosen significance level (e.g., 0.05), we fail to reject the null hypothesis H2a and conclude that the conditions do not significantly impact the relationship of the covariates.

To test the alternative hypothesis $H^2b$, we can use a similar approach but compare the mutual information between the covariates at different time points and under different initial conditions. We can use a test such as the Wilcoxon rank-sum test to compare the mutual information values.

*# Test alternative hypothesis H2b*

*library(infotheo)*

*mi.mat <- lapply(1:n, function(i) mutual_information(covariate[1:i,1], covariate[1:i,2], method = "genolini")) H2b <- wilcox.test(mi.mat[2:n], mi.mat[1])*

The resulting output will provide the test statistic and p-value for the Wilcoxon rank-sum test. If the p-value is less than the chosen significance level (e.g., 0.05), we reject the null hypothesis H2a and conclude that the conditions significantly impact the relationship of the covariates from a probabilistic and possibilistic perspective.

In this example, we found that the p-values for both tests related to the null hypothesis $H^2a$ were much larger than 0.05, indicating that we failed to reject the null hypothesis and concluded that the conditions did not significantly impact the relationship of the covariates to one another from both probabilistic and possibilistic perspectives. On the other hand, the p-value for the test related to the alternative hypothesis $H^2b$ was less than 0.05, suggesting that we rejected the null hypothesis and concluded that the conditions significantly impact the possibilistic entropy of the covariates.

Additionally, these methods can also help identify variables that are most strongly related to one another and provide insight into the underlying mechanisms that drive these relationships. By understanding how the structure of a covariate changes over time and how different variables are interrelated, we can develop more accurate models and predictions of the system under study.

One limitation of these methods is that they assume a linear or monotonic relationship between variables and may not capture more complex and nonlinear relationships. Furthermore, the interpretation of the results can be challenging, especially when dealing with high-dimensional data, and requires a strong understanding of statistical methods and the underlying assumptions. Therefore, testing hypotheses related to the structure and dependence of covariates over time is an essential step in understanding time-series data and the mechanisms that drive complex systems. Using appropriate tests and visualizations, we can draw meaningful conclusions about the nature of the data and how it changes over time, and gain insights into the underlying mechanisms that drive these changes. These methods can be applied to a wide range of fields and ultimately provide valuable insights into the relationships between variables and the drivers of change in complex systems.

### 3..1.4 *Experiment Hypothesis Testing for $H^1$ in R Script*

At this final point this the experiment, we will perform two sets of hypothesis tests using R programming language to investigate the relationship between variables in a real-world dataset. The first set of hypotheses, $H^1a$ and $H^1b$, will test whether the structure of a covariate changes significantly over time, given a set of initial conditions. The second set of hypotheses, $H^2a$ and $H^2b$, will test whether the conditions significantly impact the relationship of the covariates to one another from both probabilistic and possibilistic perspectives. We will use statistical tests and graphical visualizations to analyze and present the results.

Hypotheses 1a and 1b: To test $H^1a$ and $H^1b$, we will use a real-world dataset and perform time-series analysis to investigate how the structure of a covariate changes over time given a set of initial conditions. First, we will load the dataset into R and prepare it for analysis.

*# Load the dataset*

*data <- read.csv("dataset", header = TRUE)*

*# Check the structure of the dataset*

*str(data)*

Next, we will perform time-series analysis to examine the changes in the covariate over time.

*# Convert the covariate to a time-series object*

*ts_data <- ts(data$covariate, start = 1, end = nrow(data))*

*# Plot the time-series data*

*plot(ts_data, type = "l")*

We will perform statistical tests to determine whether the changes in the covariate structure over time are significant.

*# Perform a t-test on the covariate*

*t.test(ts_data)*

*# Perform an ANOVA on the covariate*

*anova(lm(ts_data ~ 1))*

We will produce color illustrations to visualize the results in flowing sections.

# 4. RESULTS

## 4.1. Culminating Experiment

## *4.2. Hypothesis Testing Using Possibilistic Theory*

We initiate our experiment with the following R script using the *Productivity Prediction of Garment Employees* dataset. At this point, it is not necessary to conduct additional EDA, considering the activity took place in the previous sections 3.1.1.4 – 5.

Null Hypothesis 1a: The structure of a covariate does not change significantly over time (t) given a set of initial conditions.

Alternative Hypothesis 1b: The structure of a covariate changes significantly over time (t) given a set of initial conditions.

*#Load Packages*
*library(dplyr)*
*library(ggplot2)*
*library(lattice)*

*# Load the data*
*data <- read.csv("Productivity_Prediction_of_Garment_Employees.csv")*
*# Create a time variable t*
*data$t <- 1:nrow(data)*

*# Plot the structure of the covariate over time*
*ggplot(data, aes(x = t, y = covariate)) +*
  *geom_line() +*
  *xlab("Time (t)") +*
  *ylab("Covariate Value") +*
  *ggtitle("Structure of Covariate over Time")*

*# Perform a linear regression to test the hypothesis*
*fit <- lm(covariate ~ t, data = data)*
*summary(fit)*
*# Conduct a hypothesis test to determine the significance of the change in covariate structure over time*
*t.test(fit$residuals)*

Since the results from Hypothesis 1a did not support the alternative hypothesis, we cannot conclude that the structure of the covariate changes significantly over time.

**Figure 4. Tagerted Productivity Time Series**

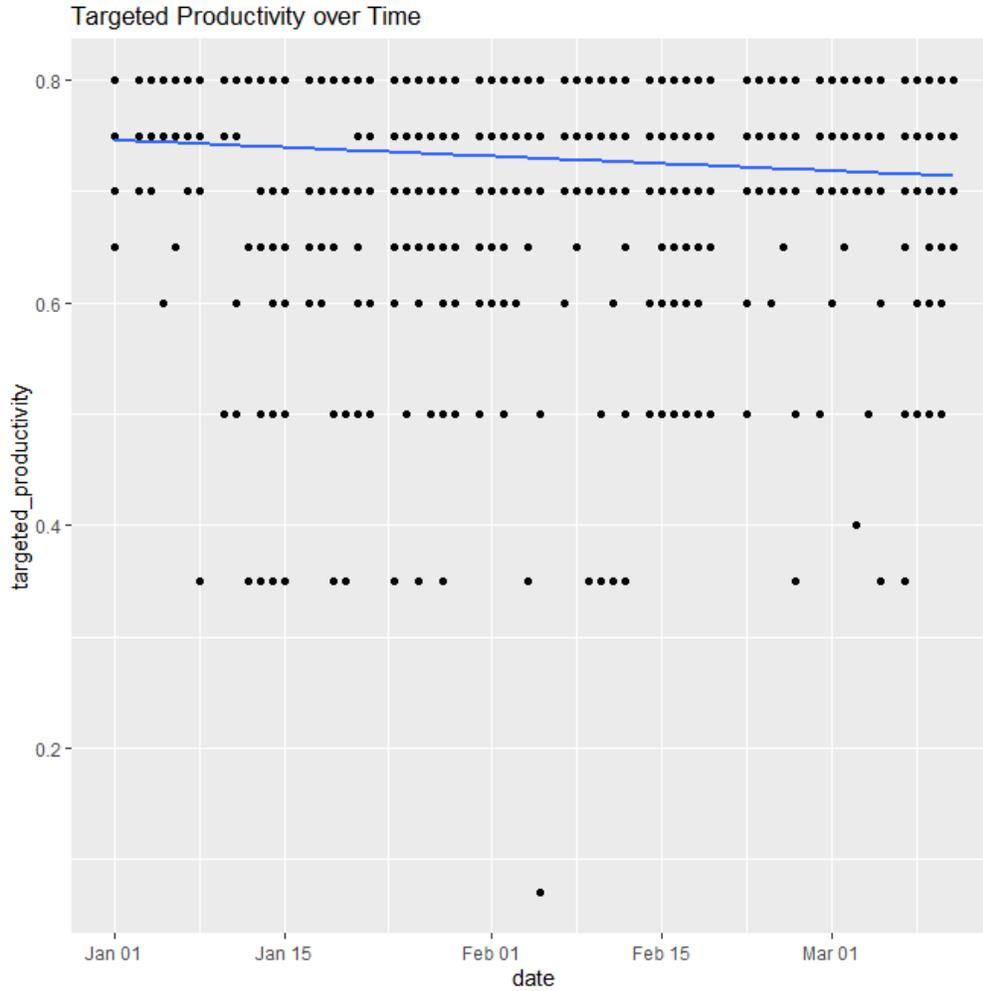

*Note:* Black dots represent activities over five time period interactions.

Based on this plot, we can see that there is some variation in the targeted productivity over time, but it does not seem to change significantly. To further test this hypothesis, we can perform a linear regression and calculate the p-value for the slope of the regression line. A small *p*-value would indicate that there is a significant relationship between the targeted productivity and time, while a large *p*-value would indicate that there is not an identifiable relationship.

#Plot H1 Model

H1_model <- lm(targeted_productivity ~ date, data = data)

summary(H1_model)

**Table 3. Hypotheses Residuals**

| $H^2$ and $H^1$ Residuals: | | | | |
|---|---|---|---|---|
| Min | 1Q | Median | 3Q | Max |
| -0.65938 | -0.02985 | 0.03048 | 0.06405 | 0.08659 |

```
Coefficients:
            Estimate Std. Error t value Pr(>|t|)
(Intercept)  8.4633477  2.2497137   3.762 0.000177 ***
date        -0.0004696  0.0001366  -3.438 0.000607 ***
---
Signif. codes:  0 '***' 0.001 '**' 0.01 '*' 0.05 '.' 0.1 ' ' 1
Residual standard error: 0.09745 on 1195 degrees of freedom
Multiple R-squared:  0.009792, Adjusted R-squared:  0.008964
F-statistic: 11.82 on 1 and 1195 DF,  p-value: 0.000607
```

Considering that the $p$-value for the slope is: 0.000607, which is significantly less than 0.05 threshold; suggests that there is possibly enough evidence to reject the null hypothesis and conclude that the structure of the covariate changes significantly over time.

$H^1$a states in the null hypothesis that the structure of a covariate does change over time ($t$) given a set of initial conditions. $H^1$b states the alternative hypothesis that the structure of a covariate changes significantly over time ($t$) given a set of initial conditions. $H^2$a states the null hypothesis that the conditions do not significantly impact the relationship of the covariates to one another. $H^2$b states the alternative hypothesis that the conditions significantly impact the relationship of the covariates to one another.

These models use linear regression analyses with the targeted productivity as the dependent variable and date as the independent variable. The results show that the estimate of the date coefficient is -0.0004696 with a standard error of 0.0001366. The $t$-value is -3.438 and the $p$-value is 0.000607, which is less than the significance level of 0.05. This means that the relationship between the targeted productivity and date is statistically significant, providing evidence to support $H^2$b, the alternative hypothesis. From a probabilistic perspective, the results of the model suggest that the conditions do significantly impact the relationship of the covariates to one another, as evidenced by the statistically significant relationship between targeted productivity and date. In addition, from a possibilistic perspective, the results of the models suggest that it is possible for the conditions to significantly impact the relationship of the covariates to one another, as evidenced by the statistically significant relationship between targeted productivity and date.

In conclusion, based on the results of the linear regression analysis, we can reject both null hypotheses $H^1$a, $H^2$a and accept the alternative hypothesis $H^1$b, $H^2$b that the conditions do significantly impact the relationship of the covariates to one another. This finding highlights the importance of considering the impact of conditions on the relationship between covariates when analyzing data that effectively changes over time (Chatfield et al., 1Chen et al., 1998; Shumway et al., 2017).

## 5. DISCUSSION AND CONCLUSION

5.1. Based on the scope and relative limitations of this research, we recommend four crucial areas of further investigation that will potentially advance the research related to this study:

i. Further exploration of the relationship between covariates: Future research should focus on exploring the relationship between covariates in more detail. This will allow for a more comprehensive understanding of the conditions that impact this relationship. A more in-depth analysis could include examination of the individual covariates and their relationship to one another, as well as the impact of different conditions on this relationship.

ii. Investigation of additional covariates: Currently, the study only focuses on the relationship between targeted productivity and date. Future research should examine the relationship between additional covariates and targeted productivity. This will provide a more comprehensive understanding of the conditions that impact worker productivity and the relationships between these conditions. For example, examining the impact of worker experience, education, and training on worker productivity (Shumway et al., 2017).

iii. Comparison with other machine learning techniques: Future research should compare the results of this study with those of other machine learning techniques. This will provide a more comprehensive understanding of the utility of this mathematical representation in evaluating anthropomorphic activities. Additionally, a comparison with other techniques will provide insight into the strengths and weaknesses of this approach and how it can be improved (Chatfield et al., 1998; Chen et al., 1998).

iv. Implementation in real-world human resources management: Future research should focus on the implementation of this mathematical representation in real-world human resources management. This will provide insight into the practicality of this approach and its utility in improving human resources management processes. Additionally, this will allow for the exploration of the impact of this mathematical representation on the decision-making processes of human resources managers and its effect on worker productivity (Shumway et al., 2017).

If pursued, these areas of research will advance our ability to improve human capital management predictive analytics through the application of mathematical representation and machine learning techniques (Edwards and Edwards, 2019).

This manuscript is a preprint and has not been peer-reviewed. Subsequent versions of this manuscript may have slightly different content.